\documentclass[epj,final]{svjour}

\usepackage{amsmath,amssymb}
\usepackage{amsfonts}
\usepackage{dsfont}
\usepackage{graphicx}

\newcommand{\ds}{\displaystyle}

\def\eqref#1{(\ref{#1})}

\newcommand{\xop}{\mathbf{b}}
\newcommand{\yop}{{\mathbf{b}^\dagger}}

\newcommand{\rop}{\mathbf{\rho}}
\newcommand{\pho}{\mathbf{\phi}}

\newcommand{\Nop}{\mathbf{n}}

\newcommand{\be}{\begin{eqnarray}}
\newcommand{\ee}{\end{eqnarray}}
\def\refeq#1{(\ref{#1})}

\def\d{\mbox d}

\def\nn{\nonumber}

\def\g{\gamma}
\def\al{\alpha}

\def\o{\omega}

\def\ve{\varepsilon}
\def\l{\left}
\def\r{\right}
\def\te{\mbox{e}}
\def\rmi{{\rm{i}}}

\def\H{{\mathcal H}}

\begin{document}
\bibliographystyle{amsplain}

\title{The $q$-deformed Bose gas: Integrability and thermodynamics}
\author{Michael Bortz \and Sergey Sergeev}
%
%
\institute{Department of Theoretical Physics,\\
         Research School of Physical Sciences and Engineering,
    Australian National University, Canberra, ACT 0200, Australia\\ \email{Michael.Bortz@anu.edu.au, Sergey.Sergeev@anu.edu.au}}

%

\abstract{
We investigate the exact solution of the $q$-deformed one-dimensional Bose gas to derive all integrals of motion and their corresponding eigenvalues. As an application, the thermodynamics is given and compared to an effective field theory at low temperatures.}

\maketitle
\section{Introduction}
One-dimensional Bethe Ansatz (BA) solvable lattice models are of intrinsic theoretical interest. On the one hand, the BA solution allows it to diagonalize the Hamiltonian and - if the number of lattice sites is $L$ - a set of further $L-1$ independent operators, the higher integrals of motion. As an application, ground-state properties as well as the thermodynamical behavior of the model may be obtained exactly. On the other hand, much insight can often be gained by formulating an effective field theory for low energies - i.e. for long-wavelength excitations. If the low-lying excitations are gapless, such a continuum theory is relativistically invariant and displays the universal properties of the model.

In this work, we investigate the one-dimensional $q$-deformed Bose model, which can be viewed as a lattice regularization of the one-dimensional Bose gas. The exact solution of this $q$-Bose model was found as a special case of a more involved $q$-deformed quantum lattice-model in \cite{bog92,bog93}, and obtained directly via the algebraic BA in \cite{bul95}. It turned out that for small coupling, the continuum limit of this model is the one-dimensional Bose gas with $\delta$-interaction, also solved by BA \cite{ll63}. This model can be cast into the form of a Luttinger liquid \cite{hal81,caz04}. The corresponding critical behaviour of the $q$-Boson model was analyzed in \cite{bog94}. Very recently \cite{che06}, the Luttinger liquid fixed point at zero temperature together with the integrability of the model were used to calculate the three-body correlation function. 

Our motivation to reconsider the $q$-Boson model is twofold. First of all, we would like to point out the diagonalization of {\em all} integrals of motion (IM) in the framework of the algebraic BA. In this approach, the IM are obtained from a generating functional, the transfer matrix. We use a slightly different, but equivalent, transfer matrix compared to the one employed by the authors cited above. This allows us to obtain rather simple expressions for the IM. Our second motivation is the study of thermodynamic quantities like the specific heat $C(T)$ and the charge susceptibility $\chi_c(T)$. Although the equation for the thermodynamical potential has already been found in \cite{bul95}, no attention has been paid yet to the actual calculation of (experimentally accessible) susceptibilities. Apart from the physical insight that these quantities offer, their low-temperature behaviour is of special interest: it reveals the operator-content of the effective field theory including irrelevant operators which necessarily appear when performing the continuum limit. Whereas it is due to the Luttinger fixed point that $\lim_{T\to 0} C(T)/T$ is constant, the leading $T$-dependent corrections to these constants are governed by the leading irrelevant operators. The same is true for a low-temperature expansion of $\chi_c(T)$, as specified in section \ref{effth}. For the $XXZ$ Heisenberg chain, these were determined in \cite{luk98} and excellent agreement with the finite-temperature magnetic susceptibility of that model was found in \cite{sir05b}. In this work, the leading irrelevant operators of the $q$-Bose model are given, without determining their amplitudes.

This article is organized as follows. In the next section, we introduce the model, summarize its exact solution and give the IM. The third section consists of the formulation of the effective field theory, whereas the fourth section is dedicated to calculating both ground-state and thermodynamic properties from the exact solution, while making contact with the field-theoretical predictions. We end with a short summary. 
\section{Integrability of the $q$-Bose model}

In this section, we introduce the model and summarize its exact
solution.

\subsection{Algebra of observables, Hamiltonian and Baxter equation.}

Let $\xop,\yop,\Nop$ be the generators\footnote{Our definition
corresponds to \cite{bog92} by $q\to q^{-1}$, $\xop\to
\sqrt{1-q^{-2}} \mathbf{B}$.} of $q$-oscillator algebra:
\be
\xop\yop=1-q^{2\Nop+2}\;,\quad \yop\xop = 1-q^{2\Nop}\;,\nn\\
\quad \xop
\Nop=(\Nop+1)\xop\;,\quad \yop \Nop=(\Nop-1)\yop\;\nn
\ee
with $0\leq q<1$. The $q$-oscillator is represented in the Fock space
$\mathcal{F}$:
\begin{equation}\label{Fock}
\Nop|n\rangle=n|n\rangle\;; \quad n = 0,1,2,3,\dots\;;\;
\xop|0\rangle =0\;.
\end{equation}
Consider a chain of length $L$ with the local $q$-oscillator
algebra assigned to each site $\ell=1,\ldots,L$. We take the quantum Lax
operator in the form
\begin{equation}\label{Lax}
\mathcal{L}_{\ell}(u)\;=\;\left(\begin{array}{cc} u & u\yop_\ell \\
\xop_\ell & 1\end{array}\right)\;,
\end{equation}
which is equivalent to the Lax operator in
\cite{bog98,bul95,che06}. The monodromy and transfer matrices are
correspondingly
\begin{equation}\label{matrices}
T(u) \;=\; \mathcal{L}_1(u) \mathcal{L}_2(u) \cdots
\mathcal{L}_L(u)
\end{equation}
and $t(u) \;=\; \textrm{Trace} \ T(u)$. The power-series expansion
of the transfer matrix gives the commutative set of integrals of
motion (IM),
 \be t(u)&=&1 + (1-q^2) u \mathcal{P}_+ +u^2\mathcal
P_+^{(2)}+u^3\mathcal P_+^{(3)}+ \cdots \label{decomposition}\\
& &+u^{L-3}\mathcal P_-^{(3)}+u^{L-2}\mathcal P_+^{(2)}+ (1-q^2)
u^{L-1} \mathcal{P}_- +\! u^L\nn\;.
 \ee
One additional integral of motion is given by the quantum
determinant of the monodromy matrix:
\begin{equation}
\textrm{$q$-det} T(u) \;=\; u^L q^{2\mathcal{N}}\;,\quad
\mathcal{N}\;=\;\sum_{\ell=1}^L \Nop_\ell\;.
\end{equation}
The eigenvalue of $\mathcal{N}$ is the particle number $N$.

The first non-trivial operators in (\ref{decomposition}) are
 \be
\mathcal{P}_+&=&\frac{1}{1-q^2}\sum_{\ell=1}^L \
\xop_{\;\ell\;}^{} \yop_{\;\ell+1\;}^{}\;,\nn\\
\mathcal{P}_-&=&\frac{1}{1-q^2} \sum_{\ell=1}^L
\xop_{\;\ell\;}^{} \yop_{\;\ell-1\;}^{}\;,\label{opers}
 \ee
where here and in the following periodic boundary conditions apply: $\ell\equiv L+\ell$. The Hamiltonian of the $q$-Bose gas is defined by
\begin{equation}\label{N}
\mathcal{H}\;=\;- \frac{1}{2} (\mathcal{P}_+ + \mathcal{P}_-)-\mu
\mathcal{N} \;,\quad
\end{equation}
where $\mu$ is the chemical potential. Furthermore, the current operator $J$ is given by $J=\frac{1}{2\rmi} \l( \mathcal P_+-\mathcal P_-\r)$ and is equally conserved.

The model can be solved by remarkably simple coordinate and algebraical
Bethe Ans\"atze \cite{bog98,bul95,che06}. The most convenient form
of the BA equations is the Baxter equation relating an
eigenvalue of the transfer-matrix and a polynomial $Q_N$ of the
power $N$, where $N$ is the number of particles,
\begin{equation}\label{TQ}
t(u)Q_N(u)\;=\; Q_N(q^2u) + u^L q^{2N} Q_N(q^{-2}u)\;.
\end{equation}
A simple derivation of the Baxter equation is given in the
appendix. The roots of $Q_N$ are the Bethe roots,
\begin{equation}\label{Q}
Q_N(u)\;=\;\prod_{a=1}^N (u-\omega_a)
\end{equation}
and the consistency condition for (\ref{TQ}) gives the Bethe Ansatz
equations
\begin{equation}\label{BA}
\omega_a^L\;=\; \prod_{b\neq a}
\frac{q\omega_a-q^{-1}\omega_b}{q^{-1}\omega_a -q\omega_b}\;\;,\;\; n=1,\ldots,N
\end{equation}
Set $\omega_a=\exp[i k_a]$, then it follows from (\ref{N},\ref{TQ})
\begin{equation}
\mathcal{H}\;=\; \sum_{a=1}^N \varepsilon(k_a)\;,\quad
\varepsilon(k)=-\mu-\cos(k)\;.
\end{equation}

\subsection{Higher Integrals of motion.}

The power-series ansatz (\ref{decomposition}) has also been
performed in \cite{bul95} to obtain the Hamiltonian. To calculate all the other IM, it is convenient to rewrite the $\mathcal
L$-operator as
 \be
 \mathcal
L_\ell(u)&=& u e_{11} + u \yop_\ell e_{12} + \xop_\ell e_{21} +
e_{22},
 \ee
where $e_{\alpha \beta}$ is a $2\times 2$-matrix with only
non-vanishing entry 1 in row $\alpha$ and column $\beta$, such
that $e_{\al\beta} e_{\g\delta}=\delta_{\beta \g} e_{\al\delta}$.
Then one finds the coefficient of $u^2$ $(u^{L-2})$ denoted as
$\mathcal P_+^{(2)}$ ($\mathcal P_-^{(2)}$)
 \be
\mathcal P_+^{(2)}&=& \sum_{\ell=1}^L \l(\xop_\ell \yop_{\ell+2} + \xop_{\ell} \yop_{\ell+1} \xop_{\ell+2} \yop_{\ell+3}\r) \nn\\
\mathcal P_-^{(2)}&=&\l[P_+^{(2)}\r]^\dagger\nn.
 \ee
Furthermore, the operators $\mathcal P_+^{(3)}$ ($\mathcal
P_-^{(3)}$) with coefficients $u^3$ ($u^{L-3}$) read
 \be
\mathcal P_+^{(3)}&=&\sum_{\ell=1}^L \l(\xop_\ell \yop_{\ell+3} +
\xop_\ell \yop_{\ell+1} \xop_{\ell+2} \yop_{\ell+4} \r.\nn\\
& &+ \l.\xop_\ell
\yop_{\ell+2} \xop_{\ell+3} \yop_{\ell+4} \r.\nn\\
& &\l.+\xop_\ell \yop_{\ell+1}
\xop_{\ell+2} \yop_{\ell+3} \xop_{\ell+4} \yop_{\ell+5}\r)\nn\\
\mathcal P_-^{(3)}&=& \l[P_+^{(3)}\r]^\dagger\nn.
 \ee
The operators $\mathcal P_+^{(\nu)}=\l[\mathcal
P_-^{(\nu)}\r]^\dagger$ for general $\nu$ are sums
 \be
\mathcal P_+^{(\nu)}&=&\sum_{\ell=1}^L \l[\sum \xop_\ell \yop_{\ell+1}
\xop_{n_2(\ell)} \yop_{m_2(\ell)} \ldots\r.\nn\\
& &\l. \xop_{n_d(\ell)}
\yop_{m_d(\ell)}\r]\label{pnu}
 \ee
where the inner sum carries over all possible choices of pairs $\xop_{n_k(\ell)}\yop_{n_k(\ell)}$ such that $\nu=m_d(\ell)-\ell +1-N_p$, with $N_p$ being the
number of $\xop\yop$-pairs.

{}From Eq.~\refeq{TQ}, the eigenvalues of the operators $\mathcal
P_+^{(\nu)}$ are readily derived, namely for $\nu=2$:
 \be \sum_{a}
\frac{\l(1-q^2\r)}{\omega_a^2} + \sum_{a\neq a'}
\frac{\l(1-q^2\r)^2}{\omega_a \omega_{a'}} \nn,
 \ee
and for $\nu=3$:
 \be
\sum_{a}  \frac{\l(1-q^2\r)}{\omega_a^3} + 2\sum_{a\neq a'}
\frac{\l(1-q^2\r)^2}{\omega_a \omega^2_{a'}}+ \sum_{a\neq a'\neq a''}
\frac{\l(1-q^2\r)^3}{\omega_a \omega_{a'}\omega_{a''}}\nn\;.
 \ee
In the $\nu$th order, one has all possible sums
 \be
\sum_{a_1\neq a_2\neq\ldots\neq a_d}
\frac{\l(1-q^2\r)^d}{\omega^{\nu_1}_{a_1} \omega^{\nu_2}_{a_2}\ldots
\omega^{\nu_d}_{a_d}}\label{highsum},
 \ee
such that $\sum_{j=1}^d\nu_j=\nu$.

\section{Limiting cases and effective field theory}
\label{effth}
In the limiting cases $q\to 1$, $q\to 0$, the Hamiltonian is written in terms of canonical Bose operators, as shown in the first part of this section. In the second part, an effective field theory valid at low energies is given.  
\subsection{Limiting cases}
To gain a first physical insight, we write the Hamiltonian in terms of bosonic operators $a_m,a^\dagger_n$ with $\l[a_m,a^\dagger_n\r]=\delta_{m,n}$. These operators act on states of $\mathcal F$ as
\be
a|n\rangle&=&\sqrt n |n-1\rangle,\;a^\dagger|n\rangle=\sqrt{n+1} |n+1\rangle,\nn\\
a^\dagger a |n\rangle &=& n|n\rangle\label{aop}.
\ee
On the other hand, 
\be
\Nop|n\rangle &=& n|n\rangle,\;\xop |n\rangle = \sqrt{1-q^{2 n}}|n-1\rangle,\nn\\
\yop |n\rangle& =& \sqrt{1-q^{2 (n+1)}}|n+1\rangle\nn.
\ee
We can thus identify $\Nop=a^\dagger a$. Let us set $q=\te^{-\eta}$ and expand
\be
\sqrt{1-q^{-\alpha}}= \sqrt{\eta \alpha} \sum_{n=0}^\infty b_n (\eta \al)^n\label{sqr1}.
\ee
Then $\xop, \yop$ is written in terms of $a,a^\dagger$:
\be
\xop=a \sqrt{2 \eta} \,\sum_{n=0}^\infty b_n (2 \eta \Nop)^n,\;\yop= \sqrt{2 \eta} \,\sum_{n=0}^\infty b_n (2 \eta \Nop)^n a^\dagger\nn,
\ee
such that
\be
\xop_\ell \yop_{\ell+1} &=& 2\eta \, \sum_{n,m}^\infty a_\ell \, b_n b_m (2\eta)^{n+m} \l( \Nop_\ell\r)^n \, \l( \Nop_{\ell+1}\r)^m a^\dagger_{\ell+1}\nn.
\ee
Following this scheme, we write down the Hamiltonian \refeq{N} explicitly for small $\eta$, i.e. $q$ near one, including the order $\mathcal O\l(\eta^2\r)$. 
\be
\H &=&  -\frac{1+\eta^2}{2} \sum_{\ell=1}^L\l( a^\dagger_\ell a_{\ell +1} + a^\dagger_{\ell} a_{\ell-1} \r) \nn\\
& &+ \frac{\eta}{4} \sum_{\ell=1}^L\l[ \Nop_\ell\l(a^\dagger_{\ell +1} a_\ell +a^\dagger_{\ell-1} a_\ell\r)\r.\nn\\
& &\l. + \l(a^\dagger_{\ell} a_{\ell-1} + a^\dagger_\ell a_{\ell+1}\r)\Nop_\ell\r] \label{hq1}\\
& &- \frac{\eta^2}{8} \l[ \frac{13}{3} \l( \Nop_\ell a^\dagger_{\ell+1} a_\ell + a^\dagger_{\ell+1} a_\ell \Nop_{\ell+1}\r) \r.\nn\\
& &\l.+ \frac{5}{3} \l( \Nop_\ell^2+\Nop_{\ell+1}^2\r) a^\dagger_{\ell+1} a_\ell + \Nop_\ell a^\dagger_{\ell+1}a_\ell \Nop_{\ell +1}+ h.c.\r]\nn.
\ee
The first term in Eq.~\refeq{hq1} is just the free hopping. The term linear in $\eta$ describes assisted hopping processes: The amplitude is enhanced by the occupation number of the place to (or from) the hopping takes place. Note that this term acts repulsively. Such a term again is present in $\mathcal O\l(\eta^2\r)$. Additionally, hopping terms assisted with the square of the occupation number of the involved lattice sites occur. The sign of the $\eta^2$-term is negative, it lowers the energy. Generally, the interaction terms in order $\eta^\nu$ involve $2\nu+2$-many operators, which can be grouped into $a^\dagger_\ell$, $a_{\ell+1}$ and $\nu$-many number operators $\Nop_{\ell}$, $\Nop_{\ell+1}$. Thus the interaction is local in each order and consists of these kinds of assisted nearest neighbour hoppings.

Let us now consider the case where $q\gtrsim 0$. Instead of \refeq{sqr1} one expands
\be
\sqrt{1-q^\al} = \sum_{n=0}^\infty c_n q^{n \al}\nn,
\ee
such that 
\be
\xop = a \, \frac{1}{\sqrt{\Nop}} \, \sum_{n=0}^\infty c_n q^{2 \Nop},\;\yop = \frac{1}{\sqrt{\Nop}} \, \l[\sum_{n=0}^\infty c_n q^{2 \Nop}\r]\, a^\dagger\nn.
\ee
The first two orders in this $q$-expansion of the Hamiltonian read
\be
\H&=& -\frac{1+q^2}{2} \sum_{\ell=1}^L  \l[\frac{1}{\sqrt{\Nop_{\ell}}} a^\dagger_{\ell} a_{\ell+1} \frac{1}{\sqrt{\Nop_{\ell+1}}}\r.\nn\\
& & \l.+\frac{1}{\sqrt{\Nop_{\ell}}} a^\dagger_{\ell} a_{\ell-1} \frac{1}{\sqrt{\Nop_{\ell-1}}}\r]\label{hstrong}.
\ee
The next term stemming from the $\mathcal P_{\pm}$-operators is of the form
\be
\frac12 \sum_{\ell=1}^L\l[\frac{q^{2 \Nop_{\ell+1}}}{\sqrt{\Nop_{\ell+1}}} a^\dagger_{\ell+1} a_\ell \frac{1}{\sqrt{\Nop_\ell}} + \frac{q^{2 \Nop_\ell}}{\sqrt{\Nop_\ell}} a^\dagger_\ell a_{\ell+1}\frac{1}{\sqrt{\Nop_{\ell+1}}}\r]\nn.
\ee
The operator \refeq{hstrong} describes a bosonic model with nearest-neighbour hopping where the hopping amplitude is inversely proportional to the square-roots of the occupancy of the lattice sites involved. This means that the hopping amplitude equals one regardless of the occupancy. 

\subsection{The effective field theory}
It is instructive to consider the Hamiltonian \refeq{N} within the density phase representation of the operators $\xop$, $\yop$. We define operators $E^{(\pm)}$ by \cite{car68,lyn95}  
\be
E^{(+)}:=\sum_{n=0}^\infty|n+1\rangle\langle n|\,,\; E^{(-)}:=\sum_{n=0}^\infty|n\rangle\langle n+1| \nn,
\ee
such that $a^\dagger=E^{(+)}  \sqrt{\Nop+1} ,\, a=  \sqrt{\Nop+1} E^{(-)}$ and 
\be
\l[\Nop,E^{(\pm)}\r] = \pm E^{(\pm)}\nn\;.
\ee
Note that $(E^{(\pm)})^\dagger=E^{(\mp)}$. However, the $E^{(\pm)}$ are not unitary:
\be
E^{(-)}E^{(+)} = \mathds 1\,,\; E^{(+)} E^{(-)}=\mathds 1-|0\rangle \langle 0|\label{epm}.
\ee
Then
\be
\xop =  \sqrt{1-q^{2 (\Nop+1)}} E^{(-)},\, \yop = E^{(+)}\sqrt{1-q^{2(\Nop+1)}} \nn .
\ee

Let us now formulate an effective field theory at small energies, i.e. at long wavelengths and small $\eta$. Therefore we pass to the continuum via
\be
x&=& l\cdot \Delta,\, \Nop_l = \Delta\cdot \rop(x) ,\nn\\
 \l[ \rop(x),E_\pm(y)\r] &=& \pm \delta(x-y)E^{(\pm)}(y)\label{commut}.
\ee
In order to be able to treat $E_\pm$ as unitary operators, we set 
\be
\rop (x) = n + \pho(x)\nn
\ee
with $\langle \pho(x)\rangle \ll n$ and $\langle \partial_x \pho(x)\rangle \ll 1$, such that $\pho$ encodes the small-amplitude, long-wavelength-oscillations of the density around its equilibrium value $n$. Here $n$ should not be confused with the label of the Fock states. Thus we neglect the vacuum component in \refeq{epm} and set $E_\pm=\te^{\mp\rmi \Pi(x)}$. Because of the commutation relation \refeq{commut}, we have $\l[\rho(x),\Pi(y)\r] = \rmi \delta(x-y)$. This leads us to
\be
\xop &=&  \sqrt{2\eta \Delta \rop}\l(1-\frac{\rop \eta \Delta}{2}\r)\te^{\rmi \Pi}+ \mathcal{O}\l(\Delta^{5/2}\r)\nn\\
\yop &=& \te^{-\rmi \Pi}\sqrt{2\eta \Delta \rop}\l(1-\frac{\rop \eta \Delta}{2}\r)+ \mathcal{O}\l(\Delta^{5/2}\r)\nn
\ee
We now expand $-\Pi(x) + \Pi(x+\Delta) = \Delta \partial_x \Pi$ and discard higher-order terms, since in the continuum limit $\Delta \to 0$ they do not contribute. We end up with
\be
\mathcal H& =& -\int \l( \rop - \frac{\Delta}{2} n \l(\partial_x \Pi\r)^2 \r.\nn\\
& & \l.- \eta \l(n^2 + 2 n \pho(x) +  \pho^2(x)\r)\r) \d x\label{fullh}.
\ee
The first and the third term yield the ground state energy density $e=-n+\eta n^2$. It will be discarded in the rest of this section. Since $\pho(x)$ is oscillatory, $\int \pho(x) \d x=0$. Setting $\eta = c \Delta$, we end up with
\be
\mathcal H &=& \frac{\Delta}{2} \int \l[n \l( \partial_x\Pi(x)\r)^2 + 2 c \pho^2(x)\r] \d x\label{mod1}.
\ee
Before proceeding further, let us note an alternative way of carrying out the continuum limit of the Hamiltonian \refeq{hq1}: Set $a_\ell=\sqrt{\Delta} \Psi(x)$, where $x=\ell\cdot \Delta$ as in Eq.~\refeq{commut}. Then the field operators $\Psi$, $\Psi^\dagger$ obey the commutation relation $\l[\Psi(x), \Psi^\dagger(y)\r] = \delta(x-y)$. If we expand $a_{\ell+1}/\sqrt{\Delta}=\Psi(x) + \Delta \partial_x \Psi(x) + \frac{\Delta^2}{2} \partial^2_x \Psi(x)+\ldots$, and again put $\eta= c\Delta$, then after an integration by parts and within the same order as \refeq{mod1},
\be
\H&=& \frac{\Delta}{2} \int \l[ \partial_x \Psi^\dagger(x) \partial_x \Psi(x) \r.\nn\\
& & \l.+  2c \Psi^\dagger(x)\Psi^\dagger(x)\Psi(x)\Psi(x)\r]\d x \label{nls}.
\ee
The corresponding quantum mechanical $N$-particle Hamiltonian $\H_N$ reads \cite{korbook}
\be
\H_N= \frac{\Delta^2}{2} \l\{\sum_{j=1}^N\partial_{x_j}^2 + 4 c \sum_{1\leq j < k \leq N} \delta(x_j-x_k)\r\}\label{bosgas}.
\ee
In \cite{hal81,caz04} the equivalence between \refeq{mod1} and \refeq{nls} has been stated. Thus the Bose-gas in the continuum with small coupling constant presents an effective low-energy model for the $q$-deformed Bose-lattice model.

We continue the investigation of \refeq{mod1}. For the ease of notation, we set $\Delta=1$ and concentrate on the Hamiltonian density $H$, defined by $\mathcal H =: \int H \, \d x$. To bring the model \refeq{mod1} into the standard form, we define the charge velocity 
\be
v_c=\sqrt{ 2n\eta}\label{vc}
\ee
and the Luttinger parameter 
\be
K=\sqrt{\frac{n}{2\eta}}.\label{kc}
\ee
Then, by scaling $\Pi=\Pi'/\sqrt{K}$, $\pho= \sqrt{K} \pho'$, we obtain $H=\frac{v_c}{2} \l( \l(\partial_x \Pi'(x)\r)^2 + \l(\pho'(x)\r)^2\r)$, which is very closely related to the Gaussian model. We will comment on it further below. Including a chemical potential term $-\mu \sum_{\ell=1}^L \Nop_\ell$ adds a contribution $-\mu(n + \sqrt{K} \pho')$ to $H$. Upon shifting $\pho'\to \pho' + \frac{\mu \sqrt{K}}{v_c}$, we arrive at
\be
H=\frac{v_c}{2} \l( \l(\partial_x \Pi'(x)\r)^2 + \l(\pho'(x)\r)^2\r)+ \frac{\mu^2 K}{2 v_c},\label{effh}
\ee
such that for the charge susceptibility one obtains $\chi_c = \frac{K}{v_c}$. In the low-coupling limit considered here,
\be
\chi_c =  \frac{1}{2\eta}\label{chic}.
\ee
Also note that from \refeq{effh} it follows that the specific heat at low temperatures behaves as $\lim_{T\to 0} C(T)/T = \frac{\pi}{3 v_c}$. Let us emphasize again that the results (\ref{vc},\ref{kc},\ref{chic}) are valid in the limit $\eta\to 0$. Higher-order contributions can be principally obtained by pursuing the field-theoretical approach to next-leading operators. However, it is much more convenient to make use of the integrability of the Hamiltonian \refeq{N}. This will be done in the next section.

In the last part of this section, the low-temperature-asymptotics of the $q$-deformed Boson-model will be considered in the framework of the effective low-energy model. Therefore it is helpful to introduce chiral components $\phi_{R,L}$. We shortly remember how this is done for the Gaussian model, defined by
\be
H_G&=&\frac{v}{2} \l( \Pi_G^2 + (\partial_x\phi_G)^2\r),\nn\\
 \l[\phi_G(x), \Pi_G(y)\r]& =& \rmi \delta(x-y)\nn.
\ee
Here it is convenient to introduce a field $\theta_G$ such that $\Pi_G=\partial_x \theta_G$. To diagonalize the Hamiltonian and to obey the commutator, one performs a mode expansion of the fields $\phi_G,\theta_G$, \cite{luk98}. As a result, $\theta_G=\phi_L-\phi_R$, $\phi_G=\phi_L+\phi_R$, where $\phi_R$ ($\phi_L$) are the chiral components of the Bose field $\phi_G$ (i.e., the right- and left-moving parts). 

In a very similar manner, the model \refeq{effh} can be treated. Namely, the diagonalization and the commutator $\l[\phi'(x),\Pi'(y)\r]=\rmi \delta(x-y)$ lead to a mode expansion which now suggests to introduce $\theta'$ such that $\phi'=\partial_x \theta'$ and $\theta'=\phi_L+\phi_R$, $\Pi'=\phi_L - \phi_R$. From this it follows that 
\be
\frac{a}{\sqrt\Delta}& =& \Psi=\te^{\rmi(\phi_L-\phi_R)/\sqrt{K}} (n + \sqrt{K}(\phi_L + \phi_R))^{1/2}\qquad\label{psi1}\\
\frac{a^\dagger}{\sqrt\Delta} &=& \Psi^\dagger=(n + \sqrt{K}(\phi_L + \phi_R))^{1/2}\te^{-\rmi(\phi_L-\phi_R)/\sqrt{K}}. \qquad\label{psi2}
\ee
Based on these operators, the effective field theory in the continuum (that is, for low energies) is formulated. In the continuum limit, all operators that are not forbidden by symmetry are allowed - the model \refeq{effh} only constitutes the leading order. To gain insight into the low-temperature behaviour of the specific heat beyond the leading order, one has to consider next-leading contributions to \refeq{effh}. This has been done for the Heisenberg chain in \cite{luk98,sir05b}. Since the low-energy effective theory of the Heisenberg chain is a Gaussian model with next-leading terms as well, similar considerations apply in the present case. Let us enumerate the possible allowed operators. In the leading order (i.e. with scaling dimension 2) these are $\l(\partial_x \phi_{L,R}\r)^2$. Combinations of uneven numbers of $\partial_x\phi_{R,L}$ are not allowed, so there are next-leading operators of scaling dimension 4: $(\partial_x \phi_{L,R})^4, (\partial_x \phi_{L})^2 (\partial_x \phi_{R})^2,(\partial^2_x \phi_{L,R})^2,(\partial^2_x \phi_{L}) (\partial^2_x \phi_{R})$. Other relativistically invariant operators are $\cos \alpha(\phi_R+\phi_L)$ and $\cos \alpha(\phi_R-\phi_L)$. The first ones cannot appear because it is impossible to built them from \refeq{psi1}, \refeq{psi2}. The latter ones are forbidden by the $U(1)$-symmetry of the model: namely, multiplying $a$ ($a^\dagger$) by $\te^{\rmi \gamma}$ ($\te^{-\rmi \gamma}$) does not change the model. From \refeq{psi1}, \refeq{psi2} one sees that such a multiplication is equivalent to shifting $\phi_L$, $\phi_R$ in opposite directions. However, an operator $\cos \alpha (\phi_R-\phi_L)$ does not preserve this symmetry. We conclude that the next-leading operators are of scaling dimension $2 n$, $n\geq 2$. According to the calculations done for the Heisenberg chain in \cite{luk98,sir05b}, the low-temperature expansions of the specific heat and charge susceptibility then read
\be
\frac{C(T)}{T} &=& \sum_j \g_j T^{2 j},\, \g_0 = \frac{\pi}{3 v_c}\nn\\
\chi_c(T) &=& \sum_j \delta_j T^{2 j},\, \delta_0 =\frac{K}{v_c}\nn.
\ee
The constant $\delta_1$ can be calculated from the low-temperature expansion of the thermodynamical potential $g$ at constant $n$, $g(T)+\mu n=e - \frac{\pi}{6 v_c} T^2$. Since $\chi_c^{-1}(T)=\partial^2_n g(T)$, one obtains 
\be
\delta_1&=& \frac{\pi}{6} \l(\frac{K}{v_c}\r)^2 \partial^2_n v_c^{-1}\label{delta1}.
\ee
The Bethe Ansatz solution allows it to verify numerically that 
\be
C(T)/T = \frac{\pi}{3 v_c} + \g_1 T^2 \,,\; \chi_c(T)=\frac{K}{v_c} + \delta_1 T^2+\delta_2 T^4,\qquad\label{ct12}\,
\ee
with unknown coefficients $\g_1$, $\delta_2$. 

\section{Ground-state properties and thermodynamics from Bethe ansatz}
\label{gs}
This section contains two parts. In the first part, ground-state properties are considered, whereas the second part focuses on the thermodynamics.

\subsection{Ground-state properties}
To investigate the ground-state properties, we start from the BA-equations \refeq{BA}. We rewrite them using the parameterization $\o_a= \exp\l[\rmi k_a\r]$, $q=\exp\l[-\eta\r]$:
\be
\te^{\rmi k_a L} &=& -\prod_{b=1}^N \frac{\sin\l(\frac{k_a-k_b}{2} + \rmi \eta\r)}{\sin\l(\frac{k_a-k_b}{2} - \rmi \eta\r)}\,,\; a=1,\ldots,N\label{ba2}.
\ee
Using the same argument as in \cite{korbook} (chapter I.2), one can prove that $k_a\in \mathbb{R}$, $a=1,\ldots,N$. We follow the common procedure of introducing the density $\rho(k)$ of roots in the thermodynamic limit
\be
\rho(k)&=& \frac{1}{2\pi} + \frac{1}{2\pi} \int_{-B}^B \frac{\sinh 2\eta}{\cosh 2\eta - \cos(k-q)} \rho(q) \d q\label{rho}\qquad\\
\frac{N}{L}&=:&n= \int_{-B}^B \rho(k)\, \d k \label{n}\\
\frac{E}{L}&=:& e = -\int_{-B}^B \cos k \, \rho(k) \, \d k\label{e}\,.
\ee
Let us consider the limiting cases $q\to 0$ and $q\to 1$. For $q\to 0$, that is $\eta\to \infty$, 
\be
\rho(k)&=& \frac{(n+1)}{2\pi}+ \mathcal O\l(q^2 n\cos k\r),\nn\\
B&=& \frac{n\pi}{n+1}+\mathcal O\l(q^2n\r)\label{rhoq0}\\
e&=&-\frac{n+1}{\pi} \sin \frac{n\pi}{n+1}+\mathcal O\l(q^{2}n\r)\label{qto0e}
\ee
For $q=0$, the root density is constant, a property known from the free fermion gas, and the energy is bounded with respect to the density. The order $\sim q^2 n$ is consistent with the corresponding Hamiltonian \refeq{hstrong}. Note that for $q=0$, the Bethe ansatz equations \refeq{ba2} are solved explicitly. Namely, for $q=0$, these are solved by (for the ground state)
\be
k_a=\frac{\pi}{N+L} \, (-N +2 a -1), \;\; a=1,\ldots,N\nn.
\ee
This result can also be interpreted as if the particles were free (i.e. $k_a=2 \pi a/L$), and each mode has occupancy number $n+1$. The above $q=0$-results have already been found in \cite{bog93,bul95}. There it was also shown that one can solve the BA equations for $q=0$ not only for the ground state, but for general excited states \cite{bog97,bog98}. This allows it to calculate correlation functions exactly \cite{bog97,bog98}.

The opposite limit $q\to 1$, that is $\eta \to 0$, is technically more involved. The integration kernel in Eq.~\refeq{rho} becomes singular in this limit. Instead of dealing with the integral equation \refeq{rho}, it is more convenient to perform a small-$\eta$ expansion of Eqs.~\refeq{ba2}. This technique has been applied successfully to quantum gases in \cite{gau71,bat04,batff05,batbf05}. Here we find that 
\be
k_a &=& \sqrt{\frac{4 \eta}{L}} q_a\label{kap},
\ee
where the $q_a$s satisfy the equation $q_a= \sum_{b\neq a}^N \frac{1}{q_a-q_b}$.
Thus the $q_a$s are the roots of the Hermite polynomial $H_N(q)$. Inserting Eq.~\refeq{kap} into Eq.~\refeq{n} yields
\be
e=-n+\eta n^2\label{qto1e},
\ee
which agrees with the field-theoretical result for weak coupling, stated after Eq.~\refeq{fullh}. It also coincides with the mean-field approximation of the Hamiltonian \refeq{hq1} including the order linear in $\eta$. Furthermore, from the properties of the $q_a$s, the root density behaves semi-circular-like: $\rho(k)\sim \frac{1}{4\pi \eta}\sqrt{8\eta n-k^2}$. Note that essentially the same behaviour has been found for the repulsive Bose gas \cite{gau71}, again implying the continuum limit \refeq{bosgas} for $q\to 1$. 

The model \refeq{N} is conformally invariant at $0\leq q<1$ (for $q=1$, the dispersion relation is quadratic and thus conformal invariance is broken). This can be seen most easily from the BA solution: The low-lying elementary excitations are the addition or removal of a particle with momentum around the two Fermi-points $\pm p_F$ with $p_F=\frac{\pi n}{2}$ with a {\em linear} dispersion relation. The corresponding dressed energy, parametrized by the quasimomentum $k$, is
\be
\ve_0(k)&=&-\cos k -\mu+ \nn\\
& & \frac{1}{2\pi} \int_{-B}^B \frac{\sinh 2\eta}{\cosh 2\eta - \cos(k-q)} \ve_0(q) \d q\label{ve},
\ee
where $\mu$ is to be determined such that $\ve_0(B)=0$. 
The velocity $v_c$ of these charge excitations (which is identical to the sound velocity) reads
\be
v_c&=& \l.\frac{\partial \ve_0(p)}{\partial p}\r|_{p=p_F}=\l.\frac{\partial_k\ve_0(k)}{\partial_k p(k)}\r|_{k=B}= \l.\frac{\partial_k\ve_0(k)}{2\pi \rho(k)}\r|_{k=B}\nn.
\ee
After the first equality sign, the usual relation to obtain the velocity for massless excitations with momentum $p$ and energy $\ve_0$ has been used. Then $p$ and $\ve_0$ have been parametrized by the quasimomentum $k$, and finally the relation between $p$ and $k$ has been inserted. In analogy to the consideration of limiting cases in the ground-state energy, we obtain for $v_c$:
\be
v_c(n)|_{q=0}= \frac{1}{n+1}\,\sin\frac{n\pi}{n+1}\,,\; v_c(n)|_{q\to 1}=\sqrt{2\eta n}\label{velapp}.
\ee
The last equation has been obtained from the equation for the sound velocity \cite{lie63}
\be
v_c=\l[ -\frac{L}{m n} \partial_L P\r]^{1/2}\label{vsound},
\ee
where the mass $m=1$ here and the pressure $P=-\partial_L E$, with $E=eL$ the total ground state energy. Thus inserting Eq.~\refeq{qto1e} into Eq.~\refeq{vsound}, the second result in \refeq{velapp} is obtained. Note that it agrees with Eq.~\refeq{vc}, derived in the field-theoretical approach. 

The charge velocity is inversely proportional to the charge susceptibility $\chi_c$ at $T=0$. To see this, we cite the result originally derived for the one-dimensional Bose gas \cite{korbook}, but also valid here 
\be
\chi_c= \frac{Z^2}{\pi v_c},\label{chict0}
\ee
where $Z\equiv \xi(B)$ and the function $\xi(k)\equiv 2\pi \rho(k)$. The Luttinger parameter $K$ is thus obtained as 
\be
K=Z^2/\pi= v_c \chi_c\label{lutpar}.
\ee
In the special case $q=0$ one has $Z=n+1$, so that 
\be
\chi_c(q=0)=\frac{(n+1)^3}{\pi \, \sin \frac{n \pi}{n+1}}\label{chicq0},
\ee
which diverges $\sim 1/(n\pi^2)$ at $n\to 0$ and as $\sim n^4/\pi^2$ at $n\to \infty$.

\begin{figure}
\begin{center}
\includegraphics*[width=\columnwidth]{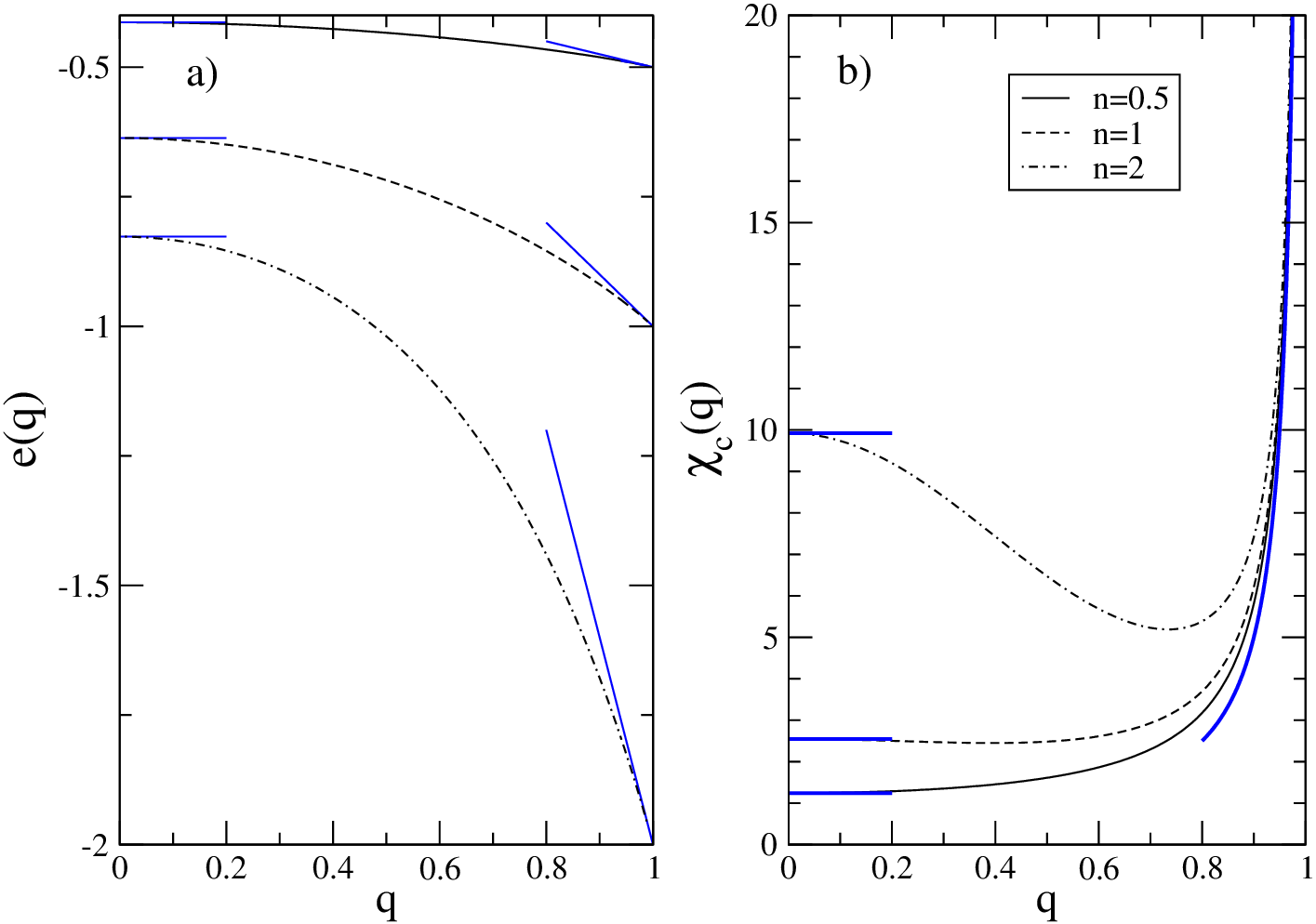}
\caption{a) The ground state energy $e(q)$ for different densities. The straight lines are the $q\to 1$ and $q\to 0$ approximations, respectively (cf. Eqs.~\refeq{qto0e}, \refeq{qto1e}; b) the charge susceptibility $\chi_c(q)$, Eq.~\refeq{chict0}. The thick blue line near $q=1$ shows the field-theoretical result \refeq{chic} which is independent of $n$. The straight lines near $q=0$ depict Eq.~\refeq{chicq0}.} 
\label{fig1}
\end{center}
\end{figure}

\begin{figure}
\begin{center}
\includegraphics*[width=\columnwidth]{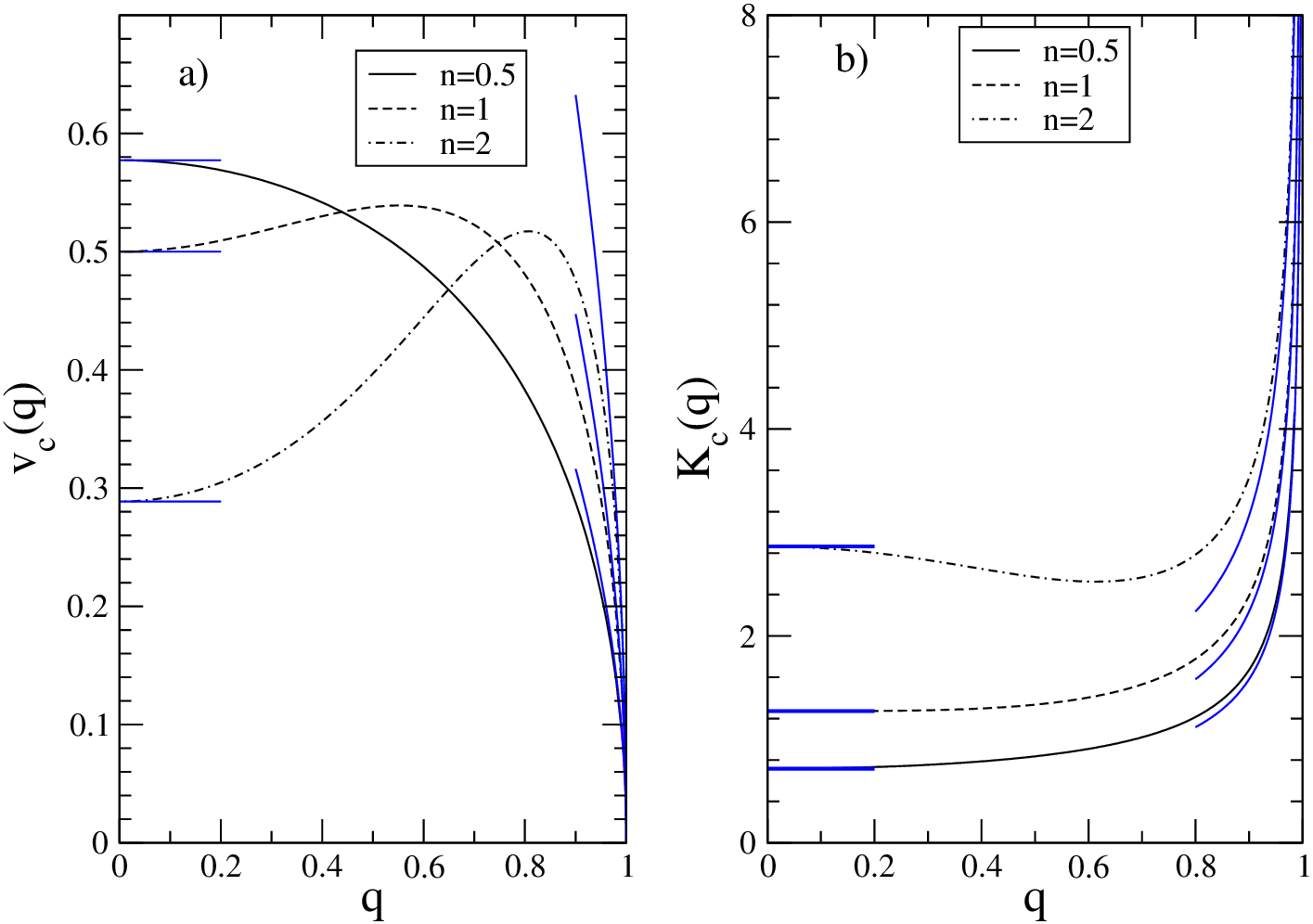}
\caption{a) The charge velocity $v_c(q)$ for different densities $n$. Additionally, near $q=1,\,0$, the analytical results \refeq{velapp} are shown; b) the Luttinger parameter $K_c$. The lines near $q=1$ denote the result \refeq{kc} from field theory, whereas near $q=0$, the values for $\l.K\r|_{q=0}=(n+1)^2/\pi$ are shown.}  
\label{fig3}
\end{center}
\end{figure}

To obtain a qualitative picture for the various quantities at arbitrary $0\leq q< 1$, Eq.~\refeq{rho}, and the derivative of Eq.~\refeq{ve} with respect to the quasimomentum are solved numerically. The results are shown in Figs.~\ref{fig1}, \ref{fig3}, together with the field-theoretical results in the weak-coupling limit $q\to 1$.
 
We finally comment on the parameters $n,\mu,B$. Since $\partial_B n = 2 \rho(B) Z>0$, the particle density is a monotonically increasing function of $B$. As $\rho(k), \ve(k)$ are $2\pi$-periodic, the integration boundary $B$ is restricted to $B\in \l[0,\pi\r]$. If $B=0$, one has $n=0$. For $B=\pi$, Eq.~\refeq{rho} together with \refeq{n} is satisfied only for $n\to \infty$. Let us now determine the corresponding $\mu$-values. Thermodynamic stability requires $\partial_\mu n>0$. As stated under Eq.~\refeq{ve}, the condition $\ve(B)=0$ fixes $\mu$. Thus $\mu=-1$ for $B=0$. If $B=\pi$, Eq.~\refeq{ve} can be treated by Fourier-transform, yielding $\mu=0$. These findings are summarized in table \ref{tab1}. In the special case $q=0$, we have from Eq.~\refeq{rhoq0} that $B=n\pi/(n+1)$, and from Eq.~\refeq{qto0e} one derives $\mu=-\partial_n e$, 
\be
\mu(n)=-\l( \frac1\pi \, \sin \frac{n\pi}{n+1} + \frac{1}{n+1} \cos\frac{n\pi}{n+1}\r)\nn.
\ee 
\begin{table}
\begin{center}
\begin{tabular}{c|cc}
$n$ & 0 & $\infty$ \\
\hline
$\mu$ & -1 & 0 \\
\hline
$B$ & 0 & $\pi$
\end{tabular}
\caption{The ranges of the parameters $n,\mu,B$ (at $T=0$).} 
\label{tab1}  
\end{center} 
\end{table}

\subsection{Thermodynamics}
As stated in the previous section, all solutions to Eqs.~\refeq{ba2} are real. This means that in order to describe equilibrium thermodynamics in the thermodynamic limit, we can follow the Yang-Yang approach \cite{yy69} which has been developed to calculate the thermodynamical potential $g(T,\mu)$ for the one-dimensional Bose gas \refeq{bosgas}. Here $T$ is the temperature and $\mu$ the chemical potential. As a result, one obtains
\be
g(T,\mu)&=& -T \int_{-\pi}^\pi \ln \l( 1+ \te^{-\ve(k)/T}\r) \,\frac{\d k}{2\pi}\nn,
\ee
where the function $\ve(k)$ is obtained from
\be
\ve(k)= -\cos k -\mu \label{vefint}\qquad\qquad\qquad\qquad\qquad\qquad\qquad\qquad\\
 -T \int_{-\pi}^\pi \frac{\sinh 2\eta}{\cosh 2\eta - \cos(k-q)}\, \ln\l(1+\te^{-\ve(q)/T}\r) \,\frac{\d q}{2\pi}\,.\nn
\ee
In the zero-temperature limit $T\to 0$, Eq.~\refeq{vefint} turns into Eq.~\refeq{ve}, with $\lim_{T\to 0} \ve = \ve_0$. 

We are interested in thermodynamic quantities like the specific heat $C(T)$ and the charge susceptibility $\chi_c(T)$ (both per lattice site) at fixed particle density $n$. Let us first consider $C(T)$. Within the grand canonical ensemble considered here, it is calculated using thermodynamic relations, cf., for example, \cite{aktj97}. From the thermodynamical potential $g(T,\mu)$, one obtains the entropy per site $S$ at fixed $\mu$ and the particle density $n$
\be
S=\partial_T g|_{\mu} \,,\qquad n=\partial_\mu g|_T\nn. 
\ee
To keep $n$ fixed, the chemical potential acquires a $T$-dependence, $\mu=\mu(T)$, such that
\be
\partial_T\mu|_n&=&-\frac{\partial_T n|_\mu}{\partial_\mu n|_T}\nn\\
\partial_T S|_n\equiv \l.\frac{S}{T}\r|_n&=& \partial_T S|_{\mu} - \frac{\l(\partial_T n|_\mu\r)^2}{\partial_\mu n|_T}\nn,
\ee
In the numerics, the partial derivatives are calculated by setting up the corresponding integral equations and iterating these, using the Fast-Fourier-Transform (FFT) algorithm. 

We have performed several consistency checks of the results following from the iteration procedure in certain limiting cases. First of all, at low temperatures the numerical data are compared with the analytical result
\be
C(T)&=& \frac{\pi}{3 v_c} \, T\label{ltc}\,,
\ee
which is derived from a saddle-point integration in \refeq{vefint} at $T\to 0$. Note that this result agrees with the prediction from conformal field theory, given that the system is conformally invariant at a finite $v_c$ (that is, at $0\leq q<1$). Furthermore, at $q=0$, 
\be
\ve(k) &=& -\cos k -\mu - T \int_{-\pi}^\pi \ln\l(1+\te^{-\ve(q)/T}\r) \,\frac{\d q}{2\pi},\nn
\ee
such that the thermodynamical potential is given by
\be
g(T,\mu)&=& -\frac{T}{2\pi} \int_{-\pi}^\pi \ln \l(1+\te^{\beta \cos k + \beta \mu - \beta g}\r)\d k\nn.
\ee
These equations are numerically easier to solve because no convolution has to be done. Especially, in the high-temperature limit $T\to \infty$ and large densities $n\gg1$
\be
g\sim T\l(-\ln n - \frac1n + \mathcal O\l(\frac{1}{n^2}\r)\r)- \frac\beta4.\label{gt}
\ee
Remarkably, the entropy $S\sim \ln n$ and the specific heat $C\sim 1/(4 T^2)$ in this limit. From the numerics for general $q>0$, we find that $C\sim \al/T^2$ with an interaction-dependent constant $\al$. 

\begin{figure}
\begin{center}
\includegraphics*[width=\columnwidth]{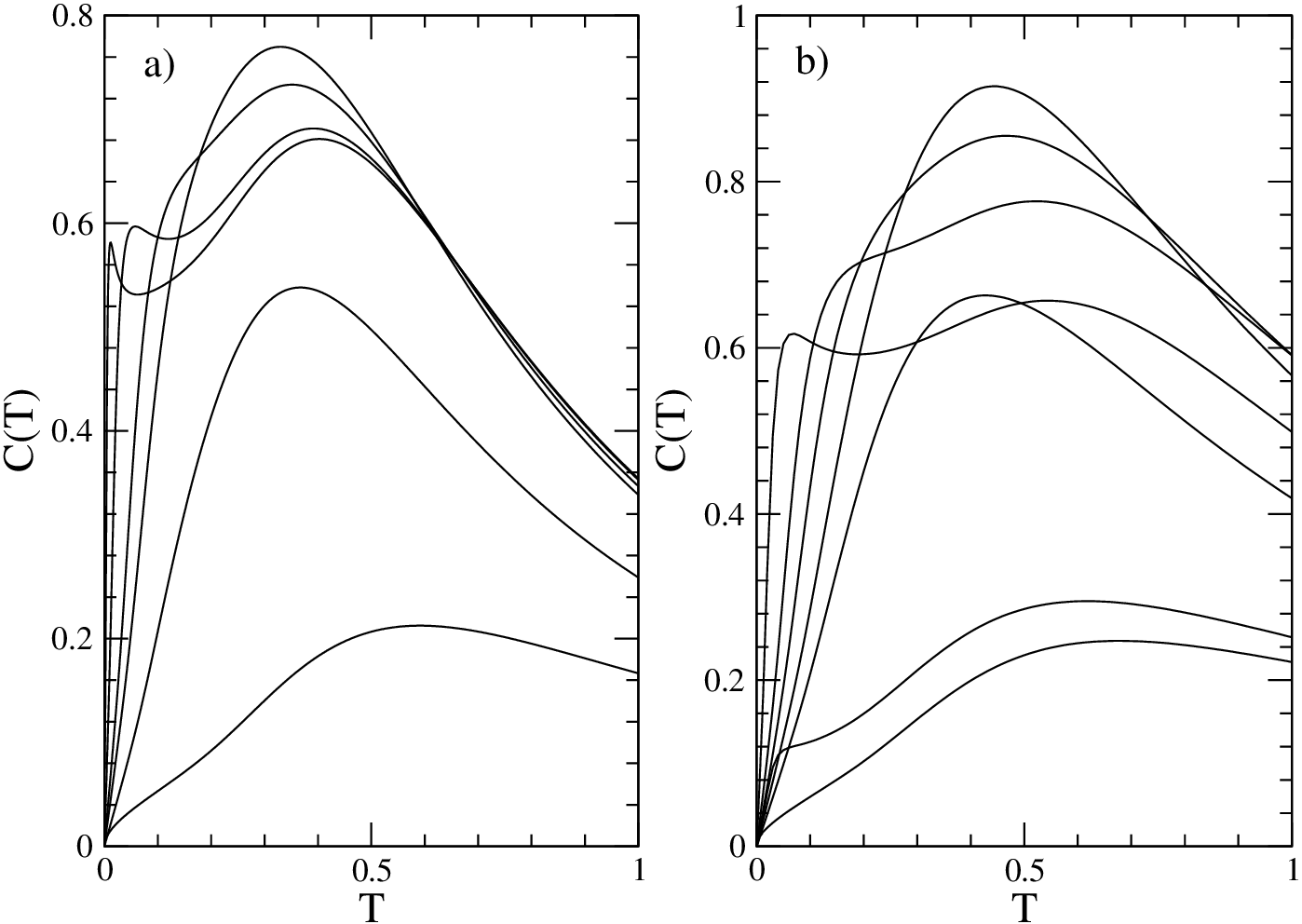}
\caption{The specific heat $C(T)$ at a) $q=0$ and b) $q=\exp\l[-0.5\r]$ for different densities (in a): $n=0.00774,\,0.649,\,2.09,\,2.95,\,5.86,\,13.9$ from bottom to top near $T=0$; in b): $n=0.01479,\,0.2306,\,1.059,\,2.356,\,3.441,\,4.658,\,8.399$ from bottom to top near $T=0$. The slope at $T\gtrsim 0$ agrees with \refeq{ltc}.}  
\label{fig4}
\end{center}
\end{figure}

\begin{figure}
\begin{center}
\includegraphics*[width=\columnwidth]{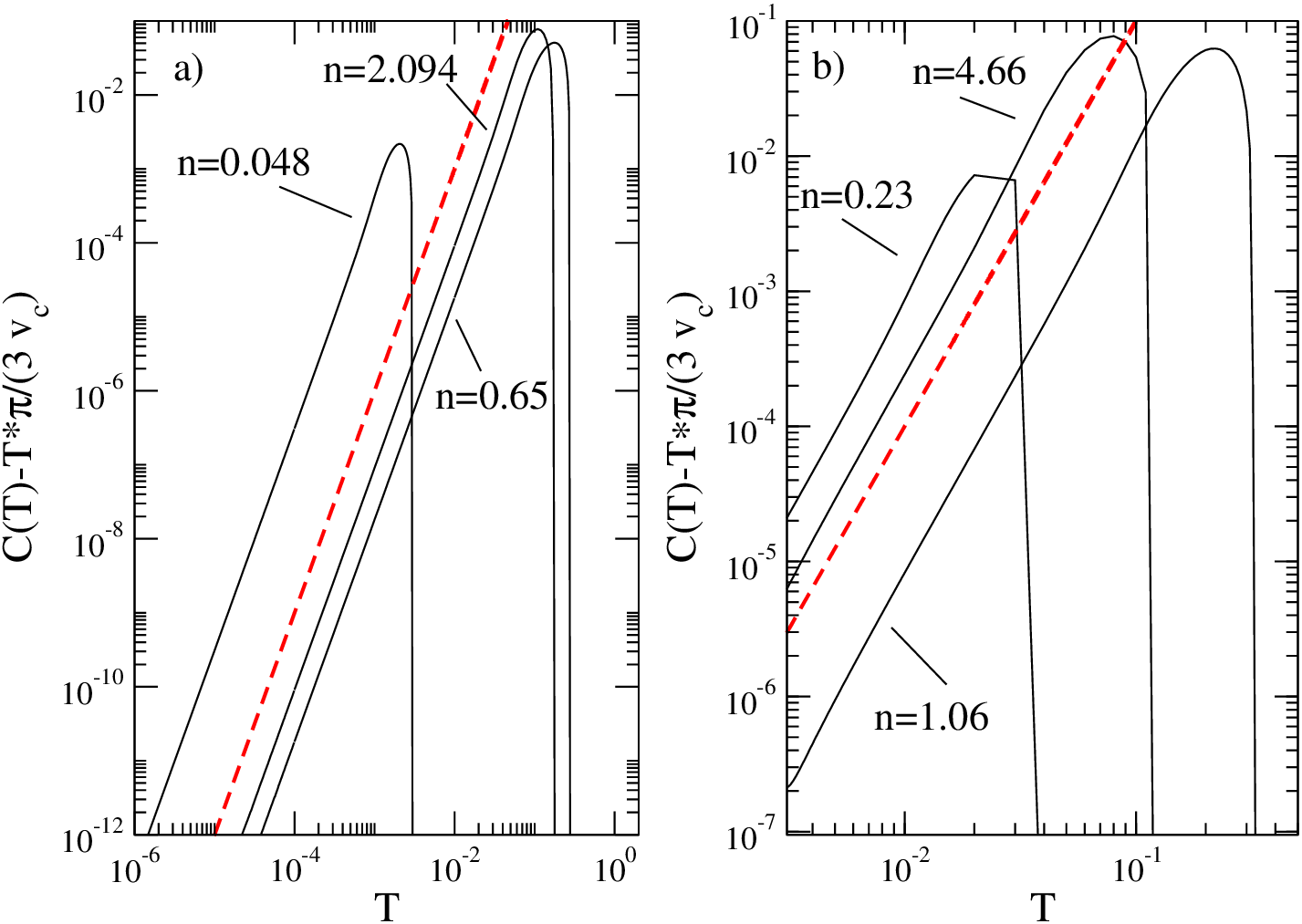}
\caption{Next-leading $T$-contribution to specific heat: $C(T)-\frac{\pi}{3v_c}\,T \propto T^3$; a) $q=0$ and b) $q=\exp\l[-0.5\r]$. For comparison, the dashed red lines are $1000\cdot T^3$ in a) ($100\cdot T^3$ in b)). This confirms that the exponent is 3 in all cases.}  
\label{fig6}
\end{center}
\end{figure}

As an illustration, Fig. \ref{fig4} shows the specific heat at the values $q=0$ and $q=\exp\l[-0.5\r]$ for different densities. It is interesting to note that at high densities, the specific heat displays a double-peak structure. We interpret this behaviour as follows. At lowest temperatures, the particle-hole excitations around the two Fermi points are activated, resulting in the slope $\propto T$ of the specific heat. At higher temperatures, all particles at each lattice site are thermally activated. The activation of this ``bulk'' causes the second peak, which is clearly discernible at high densities.   
The numerical accuracy of our data is sufficient to confirm Eq.~\refeq{ct12}. The next-leading term $\propto T^3$ in $C(T)-T \pi/(3 v_c)$ is shown in Fig.~\ref{fig6}.  

\begin{figure}
\begin{center}
\includegraphics*[width=\columnwidth]{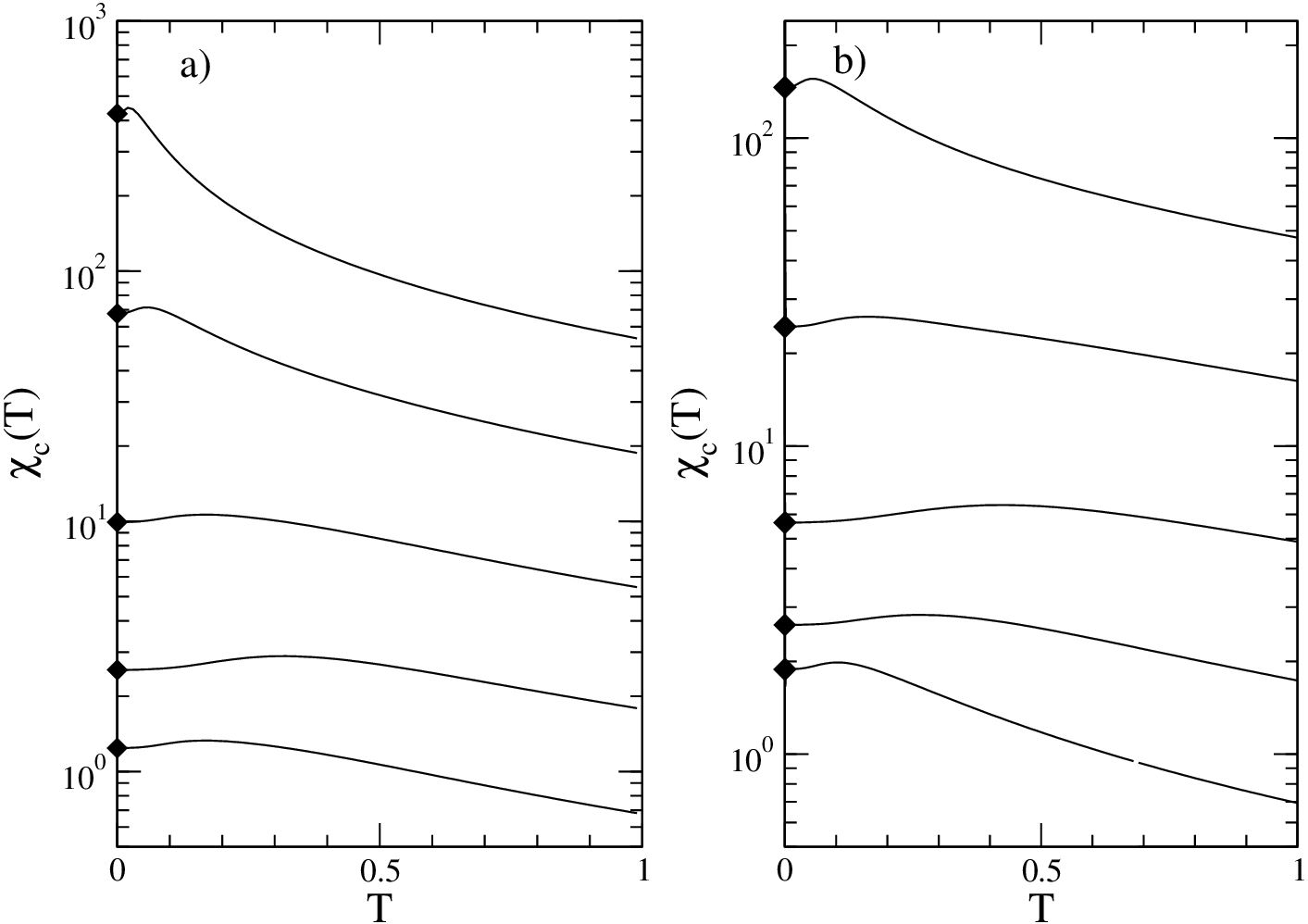}
\caption{The susceptibility $\chi_c$ at a) $q=0$ and b) $q=\exp\l[-0.5\r]$ for densities $n=0.5,\,1,\,2,\,4,\,7$ from bottom to top. Diamonds on the $y$-axis denote the $T=0
$-values obtained from \refeq{chict0}.}  
\label{fig5}
\end{center}
\end{figure}

\begin{figure}
\begin{center}
\includegraphics*[width=\columnwidth]{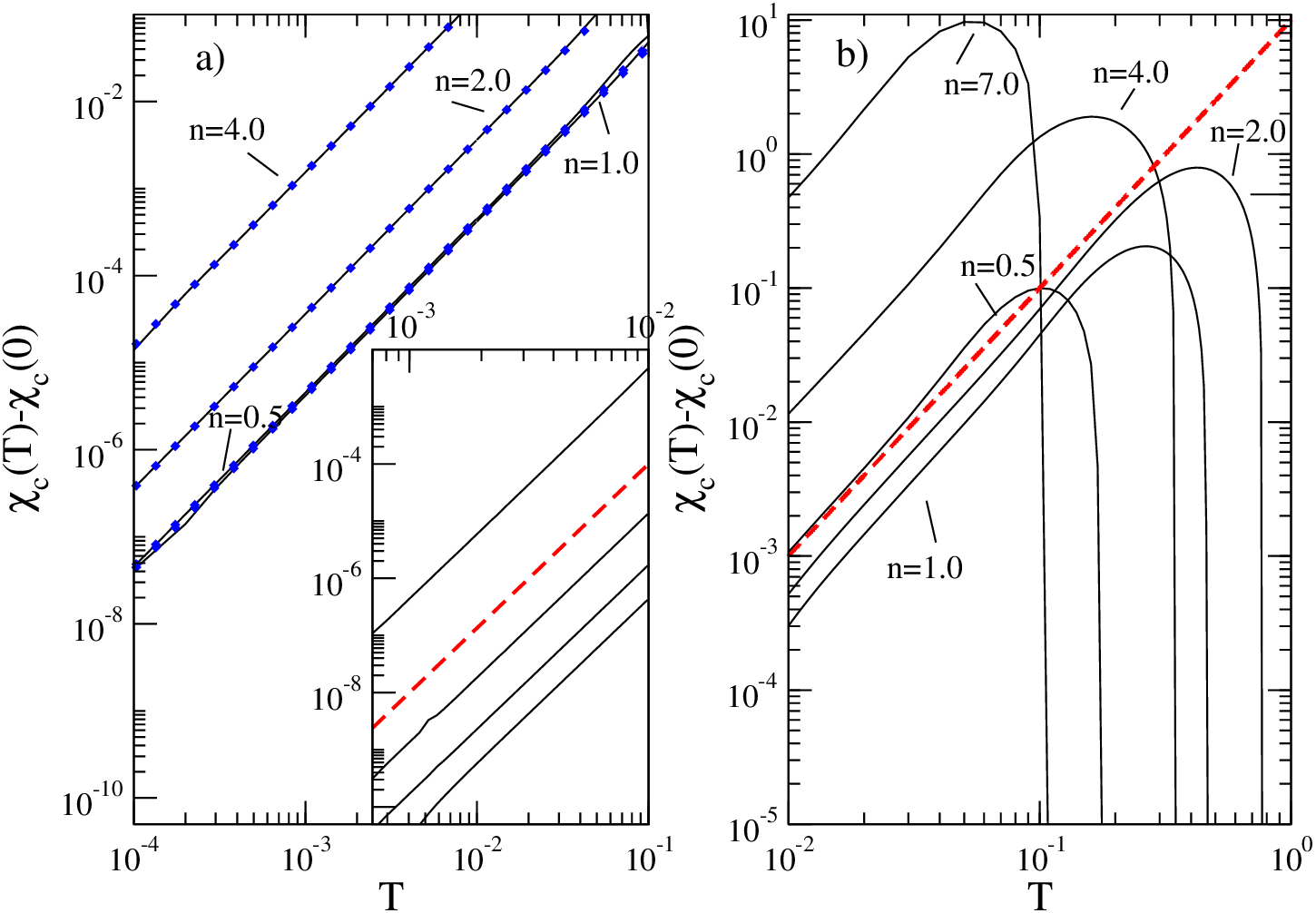}
\caption{Leading $T$-contribution to the susceptibility $\chi_c(T)-\chi_c(0)\propto T^2$ at a) $q=0$ and b) $q=\exp\l[-0.5\r]$. In a), the blue dotted lines are the exact result Eq.~\refeq{delta1}, obtained from Eqs.~\refeq{velapp}, \refeq{chicq0}. The inset shows $\chi_c(T)-\chi_c(0)-\delta_1T^2$. For comparison, the dashed red line is $10^4 \cdot T^2$ showing that the exponent is 4 in all cases. The red dashed line in b) is $10\cdot T^2$, confirming the exponent 2.}  
\label{fig7}
\end{center}
\end{figure}

We now turn our attention to the charge susceptibility $\chi_c=-\partial^2_\mu g(T,\mu)$, calculated at fixed $n$. Again for general $q$, the corresponding integral equations are solved numerically by iteration, using the FFT. The $T=0$-limit is given in Eq.~\refeq{chict0}. For $q=0$ at high temperatures, one obtains from Eq.~\refeq{gt} that $\chi_c\sim n^2/T$. In Fig.~\refeq{fig5}, the charge susceptibility is shown for $q=0$ and $q=\exp\l[-0.5\r]$ at different densities. The leading $T$-dependent contributions to $\chi_c$ are depicted in Fig.~\ref{fig7}. They confirm the field-theoretical prediction $\chi_c(T)-\chi_c(0)=\delta_1 T^2$ at low $T$. Especially, for $q=0$, the constant $\delta_1$, given in Eq.~\refeq{delta1}, is calculated explicitly from Eqs.~\refeq{velapp}, \refeq{chicq0}. This allows it to confirm numerically $\chi_c(T)-\chi_c(0)=\delta_1 T^2+\delta_2 T^4$ for $q=0$, with an unknown coefficient $\delta_2$.  

\section{Conclusion}
We investigated the exact solution of the $q$-deformed Bose gas. Emphasis was put on the higher integrals of motion by using the algebraic BA. We furthermore presented a detailed derivation of the effective field theory for low energies, with comparisons between the field-theoretical predictions and the BA. These comparisons were done both for ground-state properties and for thermodynamic quantities. Open questions include, for example, the calculation of the coefficients of the leading irrelevant operators, a high-temperature expansion of the thermodynamic quantities and the problem of formulating a quantum transfer matrix for a lattice boson model. 

\begin{acknowledgement}
\section*{Acknowledgments}
We thank X. W. Guan for pointing out Ref.~\cite{che06} to us. We also acknowledge helpful discussions with M. T. Batchelor, V. V. Bazhanov and V. V. Mangazeev. This work was supported by the Australian Research Council and by the German Research Council ({\em{Deutsche Forschungsgemeinschaft}}). 
\end{acknowledgement}
\newpage
\appendix
\section{Derivation of Baxter equation}

Consider the following systems of equations for vectors
$|\phi\rangle,|\phi'\rangle,|\phi''\rangle\in\mathcal{F}$:
\begin{equation}\label{triangular}
\begin{array}{l}
\ds ( \beta,1) \left( \begin{array}{cc} u & u\yop \\ \xop &
1\end{array}\right) \; |\phi\rangle \;=\; |\phi'\rangle \;
( \beta',1)\;,\\
\\
\ds \left( \begin{array}{cc} u & u\yop \\ \xop & 1
\end{array}\right) |\phi\rangle \left( \begin{array}{c} 1 \\
- \beta'\end{array}\right) \;=\; \left( \begin{array}{c} 1 \\
- \beta\end{array}\right) |\phi''\rangle \;.
\end{array}
\end{equation}
Solutions of these systems are
\begin{equation}
|\phi\rangle \;=\; |\phi_{ \beta, \beta'}(u)\rangle \;
\stackrel{\textrm{def}}{=} \; \sum_{n=0}^\infty  \beta^{\prime n}
c_n\left(\frac{ \beta}{ \beta'}u\right) \yop^n|0\rangle
\end{equation}
and
\begin{equation}
|\phi'\rangle  \;=\; |\phi_{ \beta, \beta'}(q^2u)\rangle \;,\quad
|\phi''\rangle \;=\; uq^{2\Nop} |\phi_{ \beta,
\beta'}(q^{-2}u)\rangle\;,
\end{equation}
where $c_n(u)$ is defined by the recurrent relation
\begin{equation}\label{reccurence}
(1-q^{2n}) c_{n}(u)\;=\; (1-u)c_{n-1}(u) +uc_{n-2}(u)
\end{equation}
with initial conditions $c_0=1$, $c_{-1}=0$. Note that $c_n(u)$ is an
$n$th power polynomial of $u$.

Let further
\begin{equation}\label{PHI}
|\Phi_{\boldsymbol{ \beta}}(u)\rangle \;=\; |\phi_{ \beta_1,
\beta_2}(u)\otimes \phi_{ \beta_2, \beta_3}(u)\otimes \cdots
\otimes \phi_{ \beta_L, \beta_1}(u)\rangle \in
\mathcal{F}^{\otimes L}\;.
\end{equation}
Since
\begin{equation}
1\;\equiv\; \left(\begin{array}{c} 0 \\ 1\end{array}\right)
(\beta_1,1)\;+\; \left(\begin{array}{c} 1 \\
-\beta_1\end{array}\right) (1,0)
\end{equation}
then
\begin{equation}
\textrm{Trace } T(u)\;\equiv \; (\beta_1,1) T(u) \left(\begin{array}{c} 0 \\
1\end{array}\right) + (1,0) T(u) \left(\begin{array}{c} 1 \\
-\beta_1\end{array}\right)\;,
\end{equation}
and it follows from (\ref{triangular}) and (\ref{PHI}),
\begin{equation}
t(u) |\Phi_{\boldsymbol{ \beta}}(u)\rangle \;=\;
|\Phi_{\boldsymbol{ \beta}}(q^2u)\rangle + u^L q^{2\mathcal{N}}
|\Phi_{\boldsymbol{ \beta}}(q^{-2}u)\rangle\;.
\end{equation}
Let further $\langle \Psi_N|$ be an eigenvector of $t(u)$, $N$
being the number of bosons. Set
\begin{equation}
Q_{\boldsymbol{ \beta},N}(u)\;=\;\langle\Psi_N|\Phi_{\boldsymbol{
\beta}}(u)\rangle \;.
\end{equation}
Since the set of $ \beta_1, \beta_2,\dots, \beta_L$ is generic,
$Q_{\boldsymbol{ \beta},N}(u)\neq 0$. The coefficients $c_n(u)$ are
polynomials, therefore $Q_{\boldsymbol{ \beta},N}(u)$ is a well
defined polynomial of $u$ of the power $N$. It satisfies the
Baxter equation (\ref{TQ}). Obviously, the set of $\boldsymbol{
\beta}$ in $Q_{\boldsymbol{ \beta},N}(u)$ corresponds to a
spectral parameter independent normalization factor:
\begin{equation}
Q_{\boldsymbol{ \beta},N}(u)\;=\;K_{\boldsymbol{ \beta}} Q_N(u)\;.
\end{equation}

\providecommand{\bysame}{\leavevmode\hbox to3em{\hrulefill}\thinspace}
\providecommand{\MR}{\relax\ifhmode\unskip\space\fi MR }
\providecommand{\MRhref}[2]{%
  \href{http://www.ams.org/mathscinet-getitem?mr=#1}{#2}
}
\providecommand{\href}[2]{#2}

\end{document}